\pdfoutput = 1
\documentclass[letterpaper]{article}
\usepackage{aaai}
\usepackage{times}
\usepackage{helvet}
\usepackage{courier}
\usepackage{amsmath}
\usepackage{graphicx}
\usepackage[colorlinks=true, allcolors=blue]{hyperref}
\usepackage{algorithm}
\usepackage[noend]{algorithmic}
\usepackage{bm}
\usepackage{amssymb}
\usepackage{amsfonts}
\usepackage{bbm}
\usepackage{amsthm}
\usepackage{subfigure}
\title{Using Drone Swarm to Stop Wildfire: A Predict-then-optimize Approach}
\author{
  Shijie Pan\textsuperscript{1,2$\dagger$}, Aoran Cheng\textsuperscript{1$\dagger$}, Yiqi Sun\textsuperscript{1}, Kai Kang\textsuperscript{3}, 
  Cristobal Pais\textsuperscript{4,5}, Yulun Zhou\textsuperscript{1$*$}, Zuo-Jun Max Shen\textsuperscript{1,5$*$} \\
  \textsuperscript{1}The University of Hong Kong, Hong Kong SAR, China \\
  \textsuperscript{2}Johns Hopkins University, Baltimore, Maryland, USA \\
  \textsuperscript{3}Shenzhen University, Shenzhen, Guangdong, China \\
  \textsuperscript{4}Amazon, Seattle, Washington, USA \\
  \textsuperscript{5}The University of California, Berkeley, Berkeley, California, USA \\
  \texttt{\{u3597137,acheng39\}@connect.hku.hk, span34@jh.edu, \{yulunzhou,maxshen,yiqisun\}@hku.hk,} \\
  \texttt{kaikang@szu.edu.cn, cpaismz@gmail.com, \{maxshen,cpaismz\}@berkeley.edu}
\thanks{$\dagger$ These authors contributed equally to this work. $*$ Corresponding authors: Yulun Zhou and Zuo-Jun Max Shen.}
}

\date{} 
\begin{document}
\maketitle
\date{} 
\maketitle
\maketitle
\begin{abstract}
\begin{quote}
Drone swarms coupled with data intelligence can be the future of wildfire fighting. However, drone swarm firefighting faces enormous challenges, such as the highly complex environmental conditions in wildfire scenes, the highly dynamic nature of wildfire spread, and the significant computational complexity of drone swarm operations. We develop a predict-then-optimize approach to address these challenges to enable effective drone swarm firefighting. First, we construct wildfire spread prediction convex neural network (Convex-NN) models based on real wildfire data. Then, we propose a mixed-integer programming (MIP) model coupled with
dynamic programming (DP) to enable efficient drone swarm task planning. We further use chance-constrained robust optimization (CCRO) to ensure robust firefighting performances under varying situations. The formulated model is solved efficiently using Benders Decomposition and Branch-and-Cut algorithms. After 75 simulated wildfire environments training, the MIP+CCRO approach shows the best performance among several testing sets, reducing movements by 37.3\% compared to the plain MIP. It also significantly outperformed the GA baseline, which often failed to fully extinguish the fire. Eventually, we will conduct real-world fire spread and quenching experiments in the next stage for further validation.

\end{quote}
\end{abstract}
\section{Introduction}
Wildfires have become progressively frequent and destructive in recent years. Numerous regions worldwide face increasingly serious wildfire threats, including loss of life, property damage, and environmental degradation. For example, in 2023, the wildfire in Nova Scotia burned over 9 million acres, significantly affecting air quality in NYC. In 2022, the United States experienced 66,255 wildfires, burning 636,031 acres; China observed 709 forest fires, resulting in 17 deaths and financial losses exceeding 2.2 billion Chinese yuan. These events underscore the urgent need for more effective wildfire management strategies \cite{lindenmayer2022adaptive}.
\begin{figure}[H] 
\centering  
\includegraphics[scale=0.55]{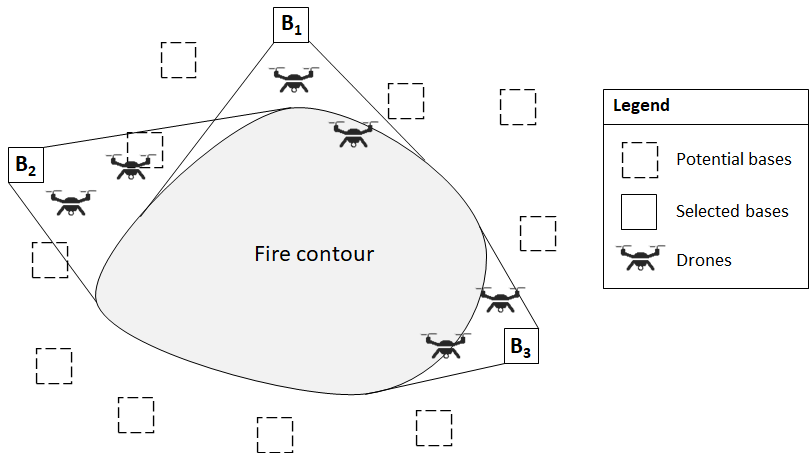}
\caption{A conceptual diagram of drone swarm firefighting.}
\label{drone swarm}
\end{figure}
\vspace{-3mm}
Drone swarm \cite{haksar2018distributed} has been recognized as an essential tool for enhancing firefighting efficiency due to its ability to target fire points precisely, and the conceptual diagram is illustrated in Figure \ref{drone swarm}. However, its practical application is not as successful as expected, due to the following problems \cite{chung2021applications,vskrinjar2019application}. First, the allocation and coordination of drone swarms often encounter difficulties in achieving the global optimum, especially in complex wildfire environments \cite{duvall2019air,wang2019vehicle}. Due to the problem's inherent complexities, existing solving algorithms can only lead to a locally optimized strategy, which is likely to occur with sub-optimal firefighting performance and thus compromises the mission's success. Current wildfire prediction models, primarily rely on simple cellular automata \cite{alexandridis2008cellular} or Markov chain models \cite{boychuk2009stochastic}, often struggle with high-precision predictions when dealing with complex variables like terrain, wind speed, and fuel moisture, leading to inefficient resource allocation during actual deployments and raising concerns about the effectiveness of emergency responses.
\vspace{-2mm}
\begin{figure}[H] 
\centering  
\includegraphics[height=3.72cm,width = 6.5cm]{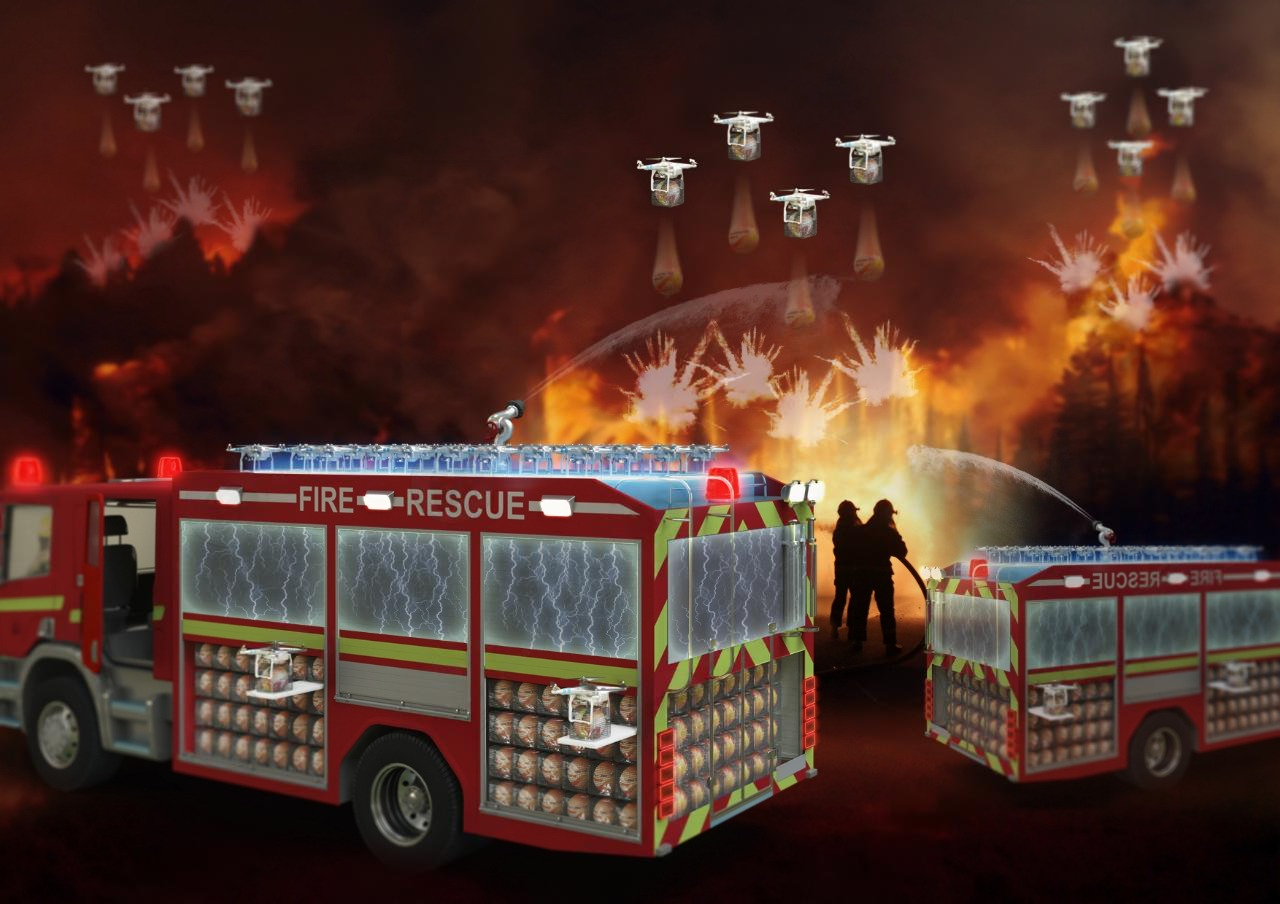}
\caption{Real-world air-drone firefighting.}
\label{drone fire}
\end{figure}
In this research, we first develop a wildfire prediction model based on Convex-NN and use it to guide the optimal drone assignment. This prediction model is grounded in data from complex wildfire environments and integrates historical data with real-time monitoring to more accurately simulate the dynamic behavior of fire spread. Compared to traditional methods \cite{johnston2008efficient}, the Convex-NN model is better equipped to handle multidimensional complexities and is closely integrated with optimization algorithms to support efficient drone swarm deployment in various environments. 

To optimize drone resource allocation strategies, we also incorporate several optimization techniques, including Mixed-Integer Programming (MIP), Dynamic Programming (DP), and Chance-Constrained Robust Optimization (CCRO), which makes our method distinct from others \cite{murray2015flying,casbeer2005forest}. This comprehensive approach ensures that the firefighting system maximizes overall benefits in diverse and complex environments, discarding inefficient decisions that favor strategies with more significant long-term advantages. Through this predict-then-optimize approach, we have improved the precision of resource allocation and significantly reduced computational burdens. Our model enables drone swarms to respond swiftly and effectively during critical moments, enhancing overall wildfire management and emergency response capabilities.
\section{Methods}
\begin{figure}[htbp]
\centering  
\includegraphics[scale=0.511]{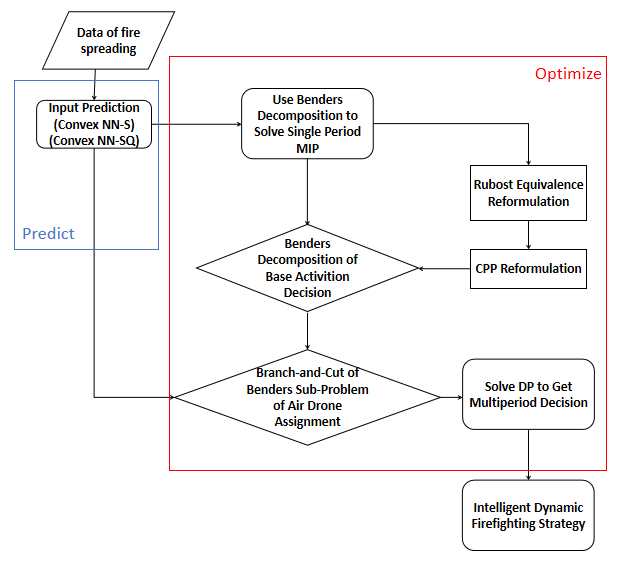}
\caption{Predict-then-optimize approach flowchart.}
\label{flowchart}
\end{figure}

This Figure \ref{flowchart} illustrates the implementation of the \textquotedblleft predict-then-optimize approach\textquotedblright. First, the dynamic behavior of wildfire spread is predicted using two models based on Convex Neural Networks: Convex-NN-S and Convex-NN-SQ. These predictions are then fed into the optimization phase, where Benders Decomposition and Branch-and-Cut are employed to solve a single-period MIP problem. The system then performs Benders Decomposition for base activation decisions (master problems) and uses Branch-and-Cut to solve subproblems related to drone task allocation. The Benders master problem is simplified through the Co-positive Programming (CPP) Reformulation for solvability. On the other hand, the chance constraint in the sub-problem is replaced by a robust equivalent form to avoid the bi-level problem. Finally, Dynamic Programming (DP) solves multi-period decisions, resulting in an intelligent dynamic firefighting strategy.
\subsection{Wildfire Spread Prediction using Convex-NN}
We develop a Convex-NN model (Convex-NN-S) to predict the spread of wildfire and a separate model (Convex-NN-SQ) is trained to make predictions under drone intervention. To effectively learn the intricate patterns of wildfire spread, we build an extensive training dataset based on real-world wildfire scenarios and the simulated data retrieved from \href{https://cell2fire.readthedocs.io/en/latest/rstfiles/WhyCell2Fire.html}{Cell2Fire} \cite{pais2021cell2fire}. Training data consists of matrices representing the state of a wildfire at a specific moment, where each matrix element could be 0 (no fire), 1 (fire burning), or 2 (fire burning with drone quenching, only used for Convex-NN-SQ training). The corresponding labels are matrices representing the subsequent state of the wildfire, capturing the effects of firefighting interventions or their absence, with each element being either 0 (no fire) or 1 (fire still burning). The training data set includes 75 environments. In the training of Convex-NN-S, the data set includes nine wildfire scenarios labeled $M_0$ through $M_8$. Each experiment involves pairwise comparisons between these datasets: specifically, $M_0$ with $M_1$, $M_1$ with $M_2$, $M_2$ with $M_3$, and so on, up to $M_7$ with $M_8$. We count 343 pairs for the 20x20 environments and 175 pairs for the 40x40 environments. Subsequently, we apply geometric transformations to augment the entire dataset, increasing the 20x20 environment pairs to 4,092 and the 40x40 environment pairs to 2,100. 

On the other hand, in the training of Convex-NN-SQ, the training data are prepared to mimic the following process. First, the fire is quenched when a drone suppresses it. Given a fire spread situation at time $t$, $\bm{\mathrm{Map}}_t$, we simulate random drone quenching, $x_{ijlt}$, to estimate fire spread after drone swarm intervention, $\bm{\mathrm{Map}}_t^{'}$. Second, the remaining fire continues to spread to its surroundings with combustibles. The spread of fire in the next time step $\bm{\mathrm{Map}}_{t+1}$ is simulated using $\bm{\mathrm{Map}}_t^{'}$ and Cell2Fire. The same process is repeated for 156 $20\times20$ grids and 100 $40\times40$ grids. The Convex-NN-SQ model is trained taking $\bm{\mathrm{Map}}_t$ and $\bm{x}_t$ (fire quenching operations) as input to predict $\bm{\mathrm{Map}}_{t+1}$ and next period burn cost $C_{t+1}$.

The predictor's convex nature allows for seamless incorporation with the Branch-and-Cut method, which optimizes task allocation among drones. 
\subsection{Optimal Drone Swarm Task Allocation}
In this paper, we make the following assumptions, which are within our expectations, to simplify the formulation and calculation. To begin with, we assume that each air drone has the same limited capacity $1$ and speed $\alpha$. The flight path of each air drone is straight and should not intersect the fire region. Battery consumption and flight time are proportional to the total distance of a certain air drone. Batteries are changed if and only if after each time slot is over and the condition of low capacity is satisfied, which means that within each length $\Delta$ period, each air drone can perform multiple round trips plus one flight towards the task point at most. To avoid large integer variables, we ignore a drone serving the same fire point multiple times in a time slot. Finally, we have one mathematical assumption about the uncertainty of \textbf{firefighting bomb-delivery} time \cite{johnston2008efficient,tymstra2010development}.
\vspace{-0.5mm}
\newtheorem{assumption}{Assumption}
\begin{assumption}
    The random time of the bomb-delivery process of each point on the map, $\tau_i,\,i=1,...,I$, follows a joint distribution $\mathbb{P}_I$ with the positive definite covariance matrix. Still, the specific distribution form remains unknown.
    \label{asump4}
\end{assumption}
\vspace{-0.5mm}
Additionally, at the beginning of each time slot $t$, we define a set containing all current fire points with size $I_t<<I$. In our new set, only the fire points are considered as $i$ to be treated, significantly reducing the feasible region of the decision variables.
\vspace{-3mm}
\paragraph{Mixed-Integer Programming.} Given the above assumptions, we develop a MIP model to optimize task allocation and route planning for drone swarm in a real-time, dynamic firefighting environment in (\ref{T1})-(\ref{C8}). In our MIP model, the decision variables use binary values to represent the activation of the base $b_{ijt}$ and the assignment of drone tasks $x_{ijlt}$.
\begin{align}
\min_{\bm{x}_t,\,\bm{b}_t}\, & \omega_1C_{t+1}+\omega_2\sum_{j}b_{jt}+2\omega_3\sum_{i,\,j,\,l}x_{ijlt}D_{ijt}\label{T1}\\
\textrm{s.t.}\, 
&G(\bm{\mathrm{Map}}_t,\,\bm{x}_{t},\,\textbf{b}_{t},\,\textbf{W}_{t})=\{\bm{\mathrm{Map}}_{t+1},\,C_{t+1}\}\label{trans}\\
&\sum_i 2x_{ijlt}D_{ijt}\leq D_lu_{lt},\,j=1,...,J,\,l=1,...,L\label{C1}\\
&x_{ijlt}\leq \min\{b_{jt},\,y_{jl},\,\pi_{ijt},\,\,z_{lt}\},\,i=1,...,I_t,\nonumber\\&\quad j=1,...,J,\,l=1,...,L\label{C2}\\
&\sum_{l,\,j} x_{ijlt}\leq \lceil g_{it}\rceil,\,i=1,...,I_t\label{C4}\\
&\mathcal{P}_{\mathbb{P}_I}\{\sum_{i,\,j} 2x_{ijlt}(D_{ijt}\alpha^{-1}+\tau_i)-\max_{i,\,j} x_{ijlt}D_{ijt}\alpha^{-1}\nonumber\\&\quad-\tau_{i'}\leq \triangle\}\geq 1-\delta,\, \forall\,\mathbb{P}_I\in\mathbb{F}_I,\,l=1,...,L\label{C6}\\
&\pi_{ijt}b_{jt}D_s\leq D_{ijt},\, i=1,...,I_t,\,j=1,...,J\label{C8}\\
& x_{ijlt}\in\{0,\,1\};\,b_{jt}\in\{0,\,1\}\nonumber
\end{align}
In (\ref{trans}), $\bm{\mathrm{Map}}_t$ is the fire intensity map at the start of time slot $t$, which includes the fire point set $\bm{g}_t$ and some not burn points. $C_{t+1}$ is the (expected) burn cost at the beginning of the next time slot $t+1$. $i'$ means $i$-label of the fire point defined by $\mathrm{argmax}_{i,\,j}x_{ijlt}D_{ij}$.

Our research focuses on three objectives: minimizing burn cost in the next period, optimizing drone battery usage, task routes, and efficient base activation operations to ensure rapid response and wildfire threat mitigation. We combine them by weights $\omega_{1},\,\omega_{2},\,\omega_3$, to ensure a rapid and efficient response and mitigation of wildfire threats.

The model inputs include the locations of fire points $\bm{\mathrm{Map}}_{t}$, the secondary fire intensity variable $g_{it}$. and drone-specific parameters such as battery capacity $u_{lt}$, period length $\Delta$, weather condition $\bm{\mathrm{W}}_t$, distance between bases and fire points $D_{ijt}$ and drone flight speed $\alpha$. The outputs are the optimal routes and task assignments for each drone. The MIP model considers various constraints, including battery limitations (\ref{C1}), drone availability (\ref{C2}), firefighting time constraint (\ref{C4}), limited operating time constraint (a chance constraint that satisfies Assumption 2 with the probability at least $1-\delta$, (\ref{C6}), and the safe distance constraint from the fire contour (\ref{C8}).

 We dynamically combine multi-phase MIP models for a multi-stage problem that includes multiple time slots $t$ and provide an efficient method to solve it in the next section. The model optimizes drone deployment to ensure firefighting operations are safe and efficient. 
 \vspace{-3mm}
\newtheorem{prop}{Proposition}
\paragraph{Myopic Dynamic Programming.} We develop a Myopic Dynamic Programming (Myopic DP) framework to minimize the burn cost and operational costs of bases and drones across multiple stages, only considering the wildfire conditions and task demands. This framework is \textquotedblleft myopic,\textquotedblright  focusing only on the costs and task allocations for the current and next stages, aligning with the wildfire predictor, which provides hourly forecasts and limits our foresight beyond the next time slot. This approach enables the model to quickly respond to dynamic wildfire changes, making optimal decisions at each time step, thereby improving drones' operational efficiency.
\begin{align*}
&R_{t+1}(\bm{\mathrm{Map}}_{t+1},\,\bm{x}_{t+1},\,\bm{b}_{t+1},\,\bm{\mathrm{W}}_{t+1},\,\bm {u}_{t+1},\,\bm{z}_{t+1})\\&=\min_{\bm{x}_{t+1},\,\bm{b}_{t+1}}\mathbb{E}_{\mathbb{P}_{I}}\{\omega_1C_{t+2}+\omega_2\sum_{j}b_{j(t+1)}\\&+2\omega_3\sum_{i,\,j,\,l}x_{ijl(t+1)}D_{ij(t+1)}+R_{t}(\bm{\mathrm{Map}}_{t},\,\bm{x}_{t},\,\bm{b}_{t},\,\bm{\mathrm{W}}_{t},\,\bm{u}_{t},\,\bm{z}_{t})\};\\
&R_T(\bm{\mathrm{Map}}_{T},\,\bm{x}_{T},\,\bm{b}_{T},\,\bm{\mathrm{W}}_{T},\,\bm{u}_{T},\,\bm{z}_{T})\\&=\min_{\bm{x}_{T},\,\bm{b}_{T}}\mathbb{E}_{\mathbb{P}_{I}}\{\omega_2\sum_{j}b_{jT}+2\omega_3\sum_{i,\,j,\,l}x_{ijlT}D_{ijT}\}.
\end{align*}
The transition constraints within the DP framework (\ref{batt1})-(\ref{state}) as follows,
\begin{align}
& u_{l(t+1)}=\upsilon_{l(t+1)}(u_{lt}-\frac{1}{D_l}\sum_{i,\,j} 
x_{ijlt}D_{ijt})+(1-\upsilon_{l(t+1)}),\nonumber\\&\quad l=1,...,L,\,t=0,...,T-1\label{batt1}\\
& \upsilon_{l(t+1)}= \lceil \frac{1}{M}\max\{0,\,u_{lt}-\frac{1}{D_l}\sum_{i,\,j} x_{ijlt}D_{ijt}-s\}\rceil,\nonumber\\&\quad l=1,...,L,\,t=0,...,T-1\label{batt2}\\
& \zeta_{l(t+1)}=\max\{0,\,\zeta_{lt}+\sum_{i,\,j} x_{ijlt}(2D_{ijt}\alpha^{-1}+\tau_{i})-\Delta\},\nonumber\\&\quad l=1,...,L,\,t=0,...,T-1\label{fly1}\\
& z_{lt}=\lceil\frac{1}{M}\zeta_{lt}\rceil,\,l=1,...,L,\,t=0,...,T-1\label{fly2}\\
&g_{iT}=\sum_{j,\,l}x_{ijlT},\,i=1,2,...,I_T;\,\bm{\mathrm{G}}_0=\bm{\mathrm{IF}};\,\bm{x}_0=\bm{0}\nonumber\\&\bm{b}_0=\bm{0};\,\bm{\mathrm{W}}_0=\bm{\mathrm{IW}};\,\bm{\zeta}_0=\bm{0};\,\bm{u}_0=\bm{D_l}\label{state}\\& 
\bm{\tau}=[\tau_{1},...,\tau_I]\sim\mathbb{P}_{I},\nonumber
\end{align}
where (\ref{batt1}) and (\ref{batt2}) ensure each drone has at least $s$ energy at the start of each time slot, which is indicated by $u_{lt}$, and preparing them for subsequent tasks. $M$ is a large positive penalty number. (\ref{fly1})-(\ref{fly2}) implies that the feasibility of the assignment of tasks is based on drone availability $z_{lt}$. (\ref{state}) illustrates the initial ($t=0$) and terminal condition ($t=T$) of DP. We can also introduce binary decision variables into the DP framework to solve scenarios where drones may return to different bases after completing tasks, allowing greater flexibility in drone operations. The formulation is shown in the Supplementary Materials.

Integrating MIP with the Myopic DP framework optimizes drone operations across multiple periods, ensuring efficient resource use and timely firefighting. Myopic DP enables real-time management of drone swarm in complex wildfire environments, significantly enhancing response the speed and overall efficiency.
\vspace{-3mm}
\paragraph{Chance-Constraint Robust Optimization.} In our MIP, we have the chance constraint (\ref{C6}) as follows,
\begin{align*}
&\mathcal{P}_{\mathbb{P}_I}\{\sum_{i,\,j} 2x_{ijlt}(D_{ijt}\alpha^{-1}+\tau_i)-\max_{i,\,j} x_{ijlt}D_{ijt}\alpha^{-1}-\tau_{i'}\\&\leq \triangle\}\geq 1-\delta,\, \forall\,\mathbb{P}_I\in\mathbb{F}_I,\,l=1,...,L,
\end{align*}
where $\mathbb{F}_I$ is a Mean-Variance (MV) robust set given by:
\begin{align*}
\mathbb{F}_I=\left\{
\mathbb{P}_I\in \mathcal{P}_0(R^I)|\,\, \mathbb{E}_{\mathbb{P}_I}[\bm{\tau}]=\bm{\mu},\,\,\mathrm{Var}_{\mathbb{P}_I}[\bm{\tau}]=\bm{\Sigma} \right\},
\end{align*}
which is a compact form for all fire points in a specific time slot  $t$. To simplify the computation complexity, we consider the $\mathbb{P}_{I_t}$ and $\mathbb{F}_{I_t}$ for only $I_t$ fire points in each time slot $t$. The robust chance constraint results in an unsolvable bi-level optimization problem with different metrics. In this paper, inspiring by \cite{liu2021time} and leveraging problem features, we find an equivalent constraint set, including a second-order positive definite constraint and several linear ones (\ref{chan1})-(\ref{chan4}) in Proposition 1, which is solvable in the Branch-and-Cut algorithm for the relaxed problem.
\begin{prop}
    In time slot $t$, the robust chance constraint (\ref{C6}) in MIP is equivalent to the following constraint sets,
    \begin{align}
    &(2\bm{x}-\bm{m})^{\top}(\frac{1-\delta}{\delta}A^{\top}\bm{\Sigma}A-(\alpha^{-1}\bm{D}+A^{\top}\bm{\mu})(\alpha^{-1}\bm{D}\nonumber\\&+A^{\top}\bm{\mu})^{\top})(2\bm{x}-\bm{m})+2\Delta(2\bm{x}-\bm{m})^{\top}(\alpha^{-1}\bm{D}+A^{\top}\bm{\mu})\nonumber\\&-\Delta^2\leq 0 \label{chan1}\\
    &\alpha^{-1}(2\bm{x}-\bm{m})^{\top}\bm{D}-\Delta+(2\bm{x}-\bm{m})^{\top}A^{\top}\bm{\mu}\leq0\label{chan2}\\
    &x_{ijlt}D_{ijt}\leq \theta_{lt},\,i=1,...,I_t,\,j=1,...,J,\,l=1,...,L\label{chan3}\\
    &\theta_{lt}\leq x_{ijlt}D_{ijt}+M(1-m_{ijlt}),\,i=1,...,I_t,\,j=1,...,J,\nonumber\\& l=1,...,L\\
    &\sum_{i,\,j} m_{ijlt}=1,\,l=1,...,L,\label{chan4}
    \end{align}
    where $\bm{x}$, $\bm{m}$ and $\bm{D}$, are vectors with dimension $I_tJ$, for decision variable $x_{ijlt}$, auxiliary variable $m_{ijlt}$ and distance between bases and fire points $D_{ijt}$, respectively. $\bm{\mu}$ is the vector of the mean of each distribution of firefighting time at fire point $i$ with dimension $I_t$. $\bm{\Sigma}$ is the covariance matrix with size $I_t\times I_t$. $M$ is a very large positive number. The matrix $A$ is a pre-conditioner to compress the dimension of the vector from $I_tJ$ to $I_t$. Specifically,
   \scalebox{0.8}{$A = \begin{bmatrix}
\bm{1} & \bm{0} & \bm{0} & \bm{0} \\
\bm{0} & \bm{1} & \bm{0} & \bm{0} \\
\bm{\vdots} & \bm{\vdots} & \bm{\ddots} & \bm{\vdots} \\
\bm{0} & \bm{0} & \bm{0} & \bm{1}
\end{bmatrix}_{I_t\times I_tJ}$},where $\bm{1}=[1,1,...,1]_{J}$ and $\bm{0}=[0,0,...,0]_{J}$.\label{CCRO}
\end{prop}
Proposition \ref{CCRO} can be proved by applying the result in \cite{ghaoui2003worst}. There exists a small $\delta$ (high confidence level) to let (\ref{chan1}) be positive definite, if $\bm{\Sigma}$ is positive definite (Assumption 2). Then, we can replace (\ref{C6}) by (\ref{chan1})-(\ref{chan4}). The bi-level part of the MIP is simplified for solving with the two-stage Benders Decomposition and Branch-and-Cut Algorithm (Algorithm \ref{alg:example}).
\subsection{A Two-Stage Algorithm: Benders Decomposition and Branch-and-Cut}
In previous sections, each single-period MIP includes a second-order positive definite constraint and a nonlinear fire contour iteration from $F_t$ to $F_{t+1}$, (\ref{trans}), making it computationally unaffordable to solve. To address this, inspired by \cite{kang2021exact,kim2018traveling,roberti2021exact}, we first introduce a Benders Decomposition to determine which base should be activated  ($b_{jt}$).
\begin{align*}
\min_{\bm{b}_t}\, & \omega_2\sum_{j}b_{jt}\nonumber+X,\\
\textrm{s.t.}\, &\pi_{ijt}b_{jt}D_s\leq D_{ijt},\, i=1,...,I_t,\,j=1,...,J\\
& b_{jt}\in\{0,\,1\},\,j=1,...,J
\end{align*}
Here, where the $X$ is the the $x_{ijlt}$-related objective that needs to be determined by cuts. Next, the decision of bases $b^*_{jt}$ can reduce the variable $x$ space, significantly lowering the computational complexity in the Benders sub-problem. The Benders master problem is a $0-1$ Mixed Integer Linear Programming (MILP). Hence, we could reformulate it into Co-Positive Programming (CPP), which is solved using an existing cutting plane method \cite{guo2021copositive}. Meanwhile, the Benders sub-problem,
\begin{align*}
\min_{\bm{x}_t}\, & \omega_1C_{t+1}+2\omega_3\sum_{i,\,j,\,l} D_{ijt}x_{ijlt}\\
\textrm{s.t.}\, &(\ref{trans})-(\ref{C4}),\,(\ref{chan1})-(\ref{chan4}),\\
& x_{ijlt}\in\{0,\,1\},\,i=1,...,I_t,\,j=1,...,J,\,l=1,...,L,
\end{align*}
is suitable to be solved by using the Branch-and-Cut algorithm since the nonlinear constraints, (\ref{trans}) and (\ref{chan1}), are convex. This method allows the relaxed sub-problem (all binary constraints are relaxed to $[0,\,1]$) to find a near-optimal solution, forming a basis for cutting plane generation. The complete version is shown in the Algorithm \ref{alg:example}.
\begin{algorithm}
   \caption{Two Stages Benders Decomposition and Branch-and-Cut Algorithm}
   \label{alg:example}
\begin{algorithmic}
\FOR{$t=1,...,T$ }
\STATE CPP reformulate Benders master problem with decision variables $b_{jt}$.
\STATE Solve Benders master problem by cutting plane method.
\STATE Define Benders subproblem based on the solution of Benders master problem, $b^*_{jt}$.
\FOR{$h=1,...,H$}
\STATE Solve the relaxed Benders subproblem, which is convex.
\STATE Check whether each current solution satisfies the constraint $0-1$. Add cutting plane constraint on the relaxed Benders sub-problem.
\ENDFOR 
\STATE Add Benders Decomposition on the Benders master problem.
\ENDFOR
\end{algorithmic}
\end{algorithm}
\section{Data and Experiments}
\subsection{Wildfire Spread Prediction Performance}
Table \ref{tab:2} summarizes the performance of the Convex-NN-S and Convex-NN-SQ predictors in 20x20 and 40x40 forest environments, the simplified real-world scenario. In wildfire spread prediction, sensitivity identifies all potential fire spread areas, specificity reduces false alarms, precision ensures accurate identification of affected regions, and accuracy evaluates the overall reliability of the predictions. As dataset complexity increases, Convex-NN-S shows a sensitivity increase from 0.8214 to 0.8995, while Convex-NN-SQ's sensitivity rises from 0.8605 to 0.9560, with both maintaining specificity above 0.9900. Convex-NN-S has the highest precision in the 20x20 environment (0.9888), which decreases with more complex datasets. Convex-NN-SQ's precision increases from 0.9528 to 0.9730. Accuracy remains consistently high across both datasets, demonstrating the predictor's robustness to varying data complexities.

\begin{figure}[htbp]
\centering  
\subfigure[20x20 forest environment]{
    \includegraphics[scale=0.18]{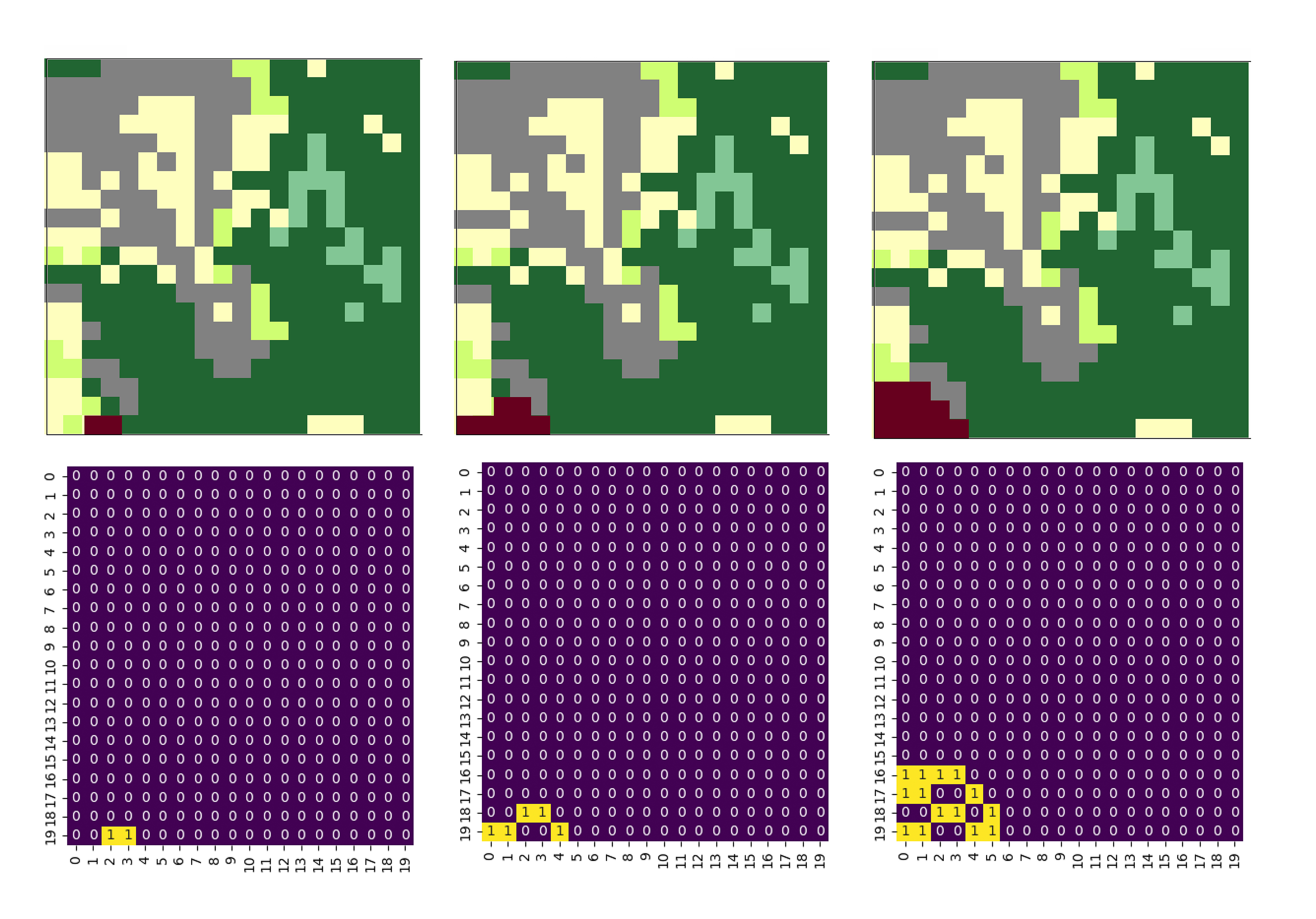}
    \label{fig:3}
}
\hspace{0.5cm} 
\subfigure[40x40 forest environment]{
    \includegraphics[scale=0.18]{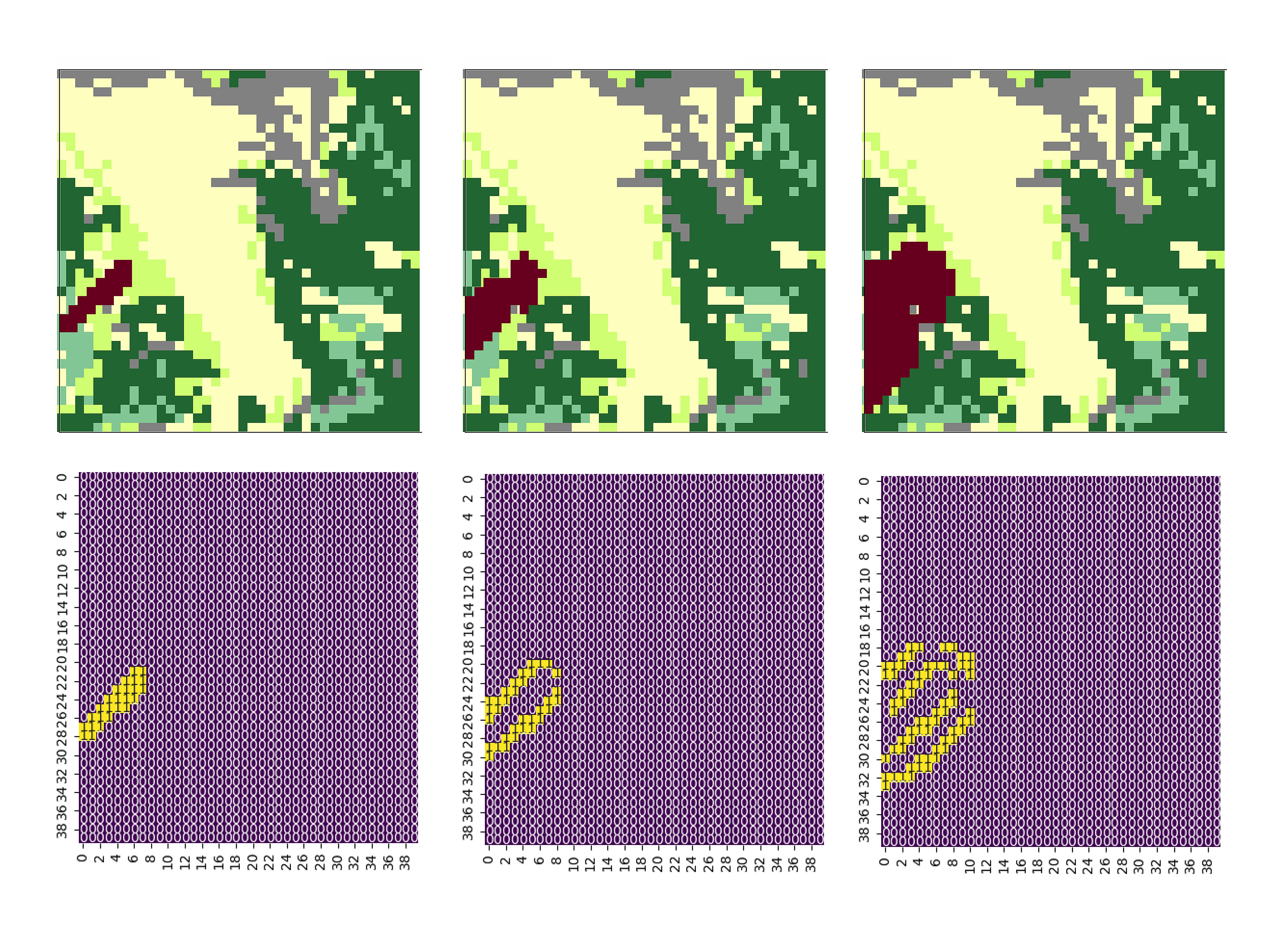}
    \label{fig:4}
}
\caption{Actual (top) and Convex-NN-S predicted (bottom) wildfire spread process for two example forest environments.}
\label{fig:combined}
\end{figure}

\begin{figure}[H]
\centering  
\subfigure{
\includegraphics[scale=0.154]{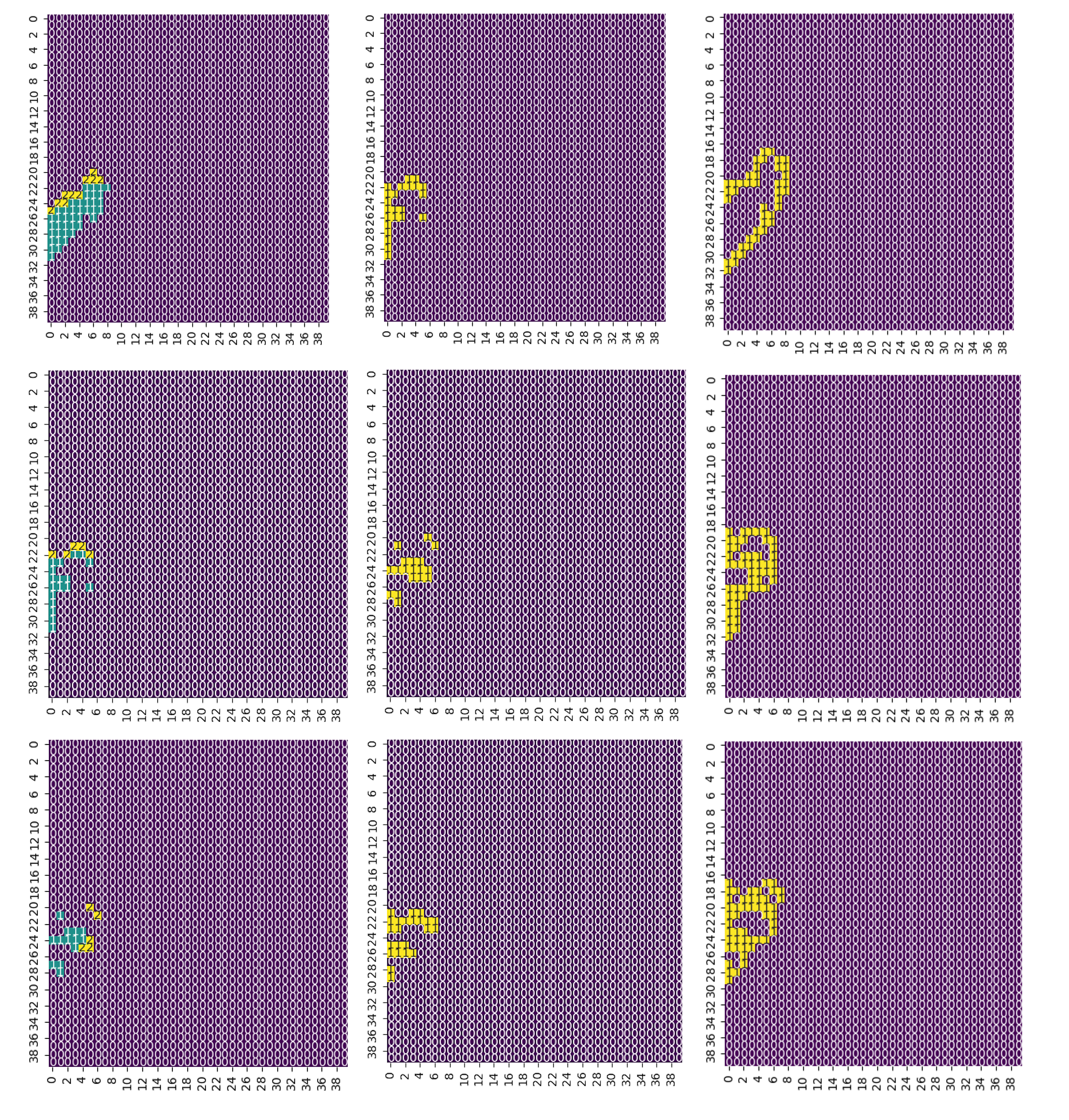}}
\caption{Three steps of Convex-NN-SQ predicted wildfire spread in a 40x40 forest with drone quenching: intervention on the left, prediction in the middle, and original spread on the right.}
\label{2}
\end{figure}

Figure \ref{fig:combined} shows the effectiveness of our fire spread predictors (Convex-NN-S) and the predictors under drone interventions (Convex-NN-SQ) are shown in Figure \ref{2}. Both predictors closely match the actual fire spread in historical datasets. Regarding the performance, the Convex-NN-SQ predictor guides our algorithm in optimizing and pre-planning drone firefighting routes, ensuring more effective fire quenching.
\subsection{Drone Swarm Quenching Algorithm Performance}
\begin{figure}[H]
\centering  
\subfigure[GA optimal drone path]{
    \includegraphics[height=4.4cm,width=6.3cm]{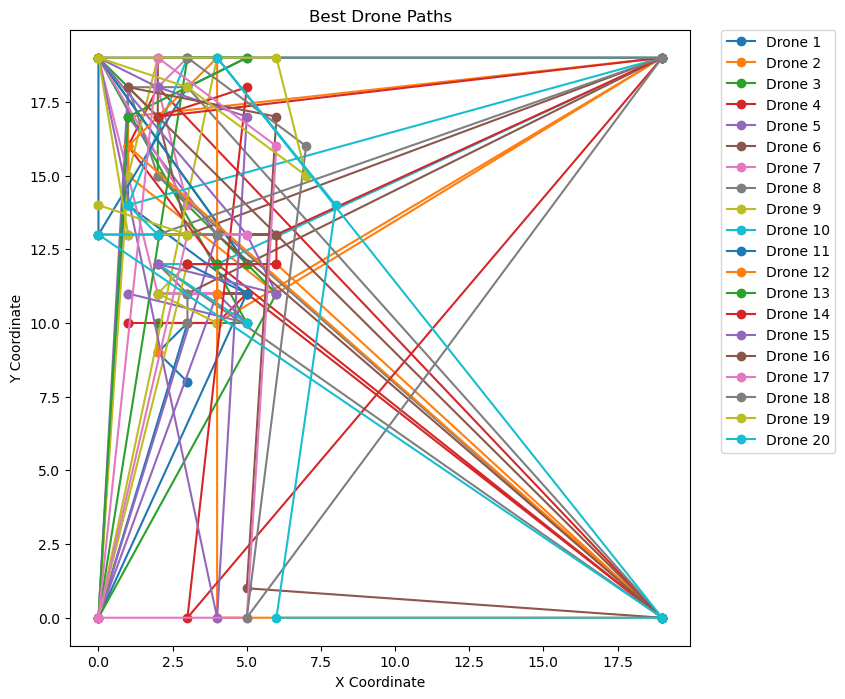}
    \label{fig:ga_path}
}
\hspace{0.5cm} 
\subfigure[MIP+CCRO optimal drone path]{
    \includegraphics[height=4.4cm,width=6.3cm]{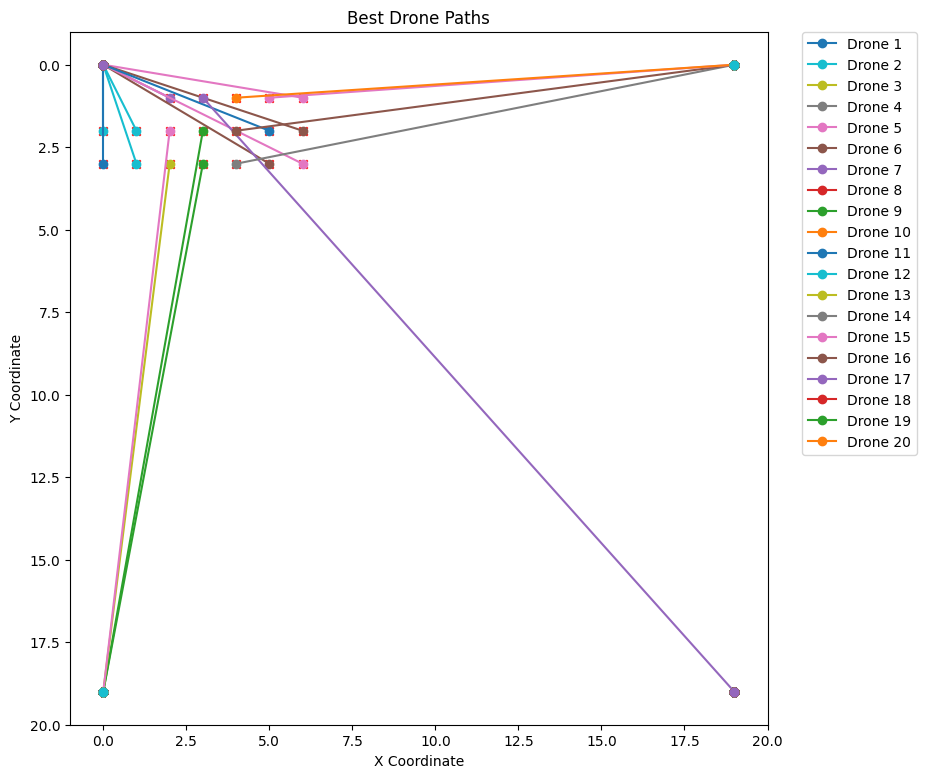}
    \label{fig:mip_ccro_path}
}
\caption{{Drone path planning for Forest 20x20-A, comparing GA at the top with MIP+CCRO at the bottom.}}
\label{fig:drone_path_comparison}
\end{figure}

We compare the model performance between the MIP+CCRO models with a Genetic Algorithm (GA), a widely adopted meta-heuristic optimization method. Our models consistently outperformed the GA in all total movements, operation rounds, and burn costs. Two metrics were conducted: the burn costs measure minimizing the fire points in the next time slot, and the other considered a sum to measure fire quenching and drone movement. The latter ensures drones efficiently reach the nearest fire points. Across four test sets, our algorithm demonstrates a clear advantage over the GA in these metrics. The wildfire predictor is applied to both the GA and MIP+CCRO, but only the MIP+CCRO successfully encodes proximity, group cooperation, and firefighting strategies, which the GA failed to understand, like Figure \ref{fig:drone_path_comparison}. These results highlight the efficiency and effectiveness of our integrated approach using Convex-NN for fire spread prediction and firefighting strategies. 

Table \ref{tab:1} compares the performance of the plain MIP, MIP+CCRO, and GA models across four scenarios. The MIP+CCRO model significantly reduces total movements compared to plain MIP, dropping movements from 308.90 to 146.95 in Environment 1 and 218.00 to 90.61 in Environment 3, dramatically reducing drones' energy consumption. The plain MIP model's major drawback is its uniform upper bound on firefighting times per period, which leads to a long waiting time for most air drones. It achieves a 37.3\% reduction in movements while still completing the firefighting task in one go, enhancing efficiency without compromising effectiveness.
\raggedbottom
\vspace{-4mm}
\begin{table*}[htbp]
  \centering
  \scalebox{0.68}{ 
  \begin{tabular}{|l|c|c|c|c|}
    \hline
    & \multicolumn{2}{c|}{\textbf{Convex-NN-S}} & \multicolumn{2}{c|}{\textbf{Convex-NN-SQ}} \\
    \hline
    & \textbf{20x20 Forest} & \textbf{40x40 Forest} & \textbf{20x20 Forest} & \textbf{40x40 Forest} \\
    \hline
    \textbf{Sensitivity} & 0.8214 & 0.8995 & 0.8605 & 0.9560 \\
    \hline
    \textbf{Specificity} & 0.9963 & 0.9900 & 0.9925 & 0.9980 \\
    \hline
    \textbf{Precision} & 0.9777 & 0.8672 & 0.9528 & 0.9730 \\
    \hline
    \textbf{Accuracy} & 0.9788 & 0.9860 & 0.9808 & 0.9950 \\
    \hline
  \end{tabular}
  }
  \caption{Performance of wildfire spread prediction models.}
  \label{tab:2}
\end{table*}
\begin{table*}[htbp]
  \centering
  \scalebox{0.715}{ 
  \begin{tabular}{|l|c|c|c|c|c|c|c|c|c|}
    \hline
    & \multicolumn{3}{c|}{\textbf{Plain MIP}} & \multicolumn{3}{c|}{\textbf{MIP+CCRO}} & \multicolumn{3}{c|}{\textbf{GA}} \\
    \hline
    & \textbf{Moves} & \textbf{Rounds} & \textbf{Burn Cost} 
                        & \textbf{Moves} & \textbf{Rounds} & \textbf{Burn Cost} 
                        & \textbf{Moves} & \textbf{Rounds} & \textbf{Burn Cost} \\
    \hline
    \textbf{Scenario 1} & 308.90 & 1 & 44 & 146.95 & 1 & 44 & 233.70 & NA & 31248 \\
    \hline
    \textbf{Scenario 2} & 23.43 & 1 & 20 & 23.43 & 1 & 20 & 46.75 & 2 & 33 \\
    \hline
    \textbf{Scenario 3} & 218.00 & 1 & 15 & 90.61 & 1 & 15 & 129.62 & NA & 31248 \\
    \hline
    \textbf{Scenario 4} & 303.91 & 1 & 64 & 187.91 & 1 & 64 & 243.55 & NA & 31248 \\
    \hline
  \end{tabular}
  }
  \caption{Comparison of different task allocation models under four distinct 20x20 forest environments.}
  \label{tab:1}
\end{table*}
\vspace{-2mm}
Although the GA model also reduces movements (233.70 in Environment 1 and 129.62 in Environment 3), it results in a much higher burn cost of 31,248, indicating a severe failure in fire containment. The MIP+CCRO model outperforms the GA model, demonstrating that the advantage lies in the MIP+CCRO approach, which optimizes drone movements while minimizing fire damage.
\vspace{-3mm}
\section{Path to Deployment}
We will deploy the predict-then-optimize approach with the Guangzhou Institute of Advanced Technology, starting with cold-state experiments to assess its feasibility. A 40x40-meter pallet filled with combustible materials will be ignited, and our algorithms will be used to quench the fire. These tests aim to optimize our algorithms, improve the accuracy of drone \textbf{firebombs}, and address uncertainties in fire quenching.
\vspace{-5mm}
A key challenge is ensuring the algorithm's accuracy and reliability, requiring ongoing optimization of the CCRO algorithm component. After these tests, we will conduct a field experiment in Jiujiang, Jiangxi, to test predict-then-optimized approach in actual wildfire conditions.
\vspace{-1mm}
\vspace{-7mm}
\section{Conclusion \& Future Work}
In this study, we propose a predict-then-optimized approach for intelligent drone swarm task allocation to stop wildfire. Specifically, we developed a wildfire predictor based on Convex-NN and integrated it with drone resources and constraints to create a CCRO-MIP, DP, and use Algorithm 1 to solve it. Our method identifies critical fire hotspots for long-term quenching, significantly improving firefighting efficiency
and reducing environmental impact. However, limitations exist, such as the assumption that \textbf{firebombs} fully quench the fire. Future work will strengthen the model, address randomness, and involve field experiments to enhance its application in natural wildfires. Ultimately, our approach improves wildfire management efficiency and public safety.
\bibliographystyle{aaai}

\end{document}